\documentclass[12pt,a4]{article}
\usepackage{amsmath}
\usepackage{bm}
\usepackage{amsfonts}
\usepackage{amssymb}
\usepackage{graphics}
\usepackage[normal]{caption2}
\usepackage{subfigure}
\usepackage{rotating}
\usepackage{citesort}
\def\be{\begin{equation}}
\def\ee{\end{equation}}
\def\ba{\begin{array}}
\def\ea{\end{array}}
\def\bea{\begin{eqnarray}}
\def\eea{\end{eqnarray}}

\begin{document}
\baselineskip 20pt \setlength\tabcolsep{2.5mm}
\renewcommand\arraystretch{1.5}
\setlength{\abovecaptionskip}{0.1cm}
\setlength{\belowcaptionskip}{0.5cm}
\pagestyle{empty}
\newpage
\pagestyle{plain} \setcounter{page}{1} \setcounter{lofdepth}{2}
\begin{center} {\large\bf Study of nuclear dynamics of neutron-rich colliding pair at energy of vanishing flow}\\
\vspace*{0.4cm}

{\bf Sakshi Gautam} \footnote{Email:~sakshigautm@gmail.com}\\
{\it  Department of Physics, Panjab University, Chandigarh -160
014, India.\\}
\end{center}

\section*{Introduction}

The collective transverse in-plane flow has been used extensively
over the past three decades to study the properties of hot and
dense nuclear matter., i.e., the nuclear matter equation of state
(EOS) and in-medium nucleon-nucleon cross section. It has been
reported to be highly sensitive to the entrance channel parameters
like incident energy, colliding geometry and system size. The
energy dependence of flow led to its disappearance at a particular
incident energy called energy of vanishing flow (EVF) or balance
energy (E$_{bal}$). A large number of theoretical studies have
been carried out in the past studying the sensitivity of E$_{bal}$
to the system size and colliding geometry.
 Role of isospin degree of freedom in
collective transverse in-plane flow and its disappearance has also
been a matter of great interest for the past decade \cite{pak97}.
The availability of radioactive ion beams (RIBs) around the world
helps in carrying out the studies on the matter lying far away
from the stability line. A number of studies have been carried out
in the recent past to see the role of isospin degree of freedom in
collective flow and its disappearance \cite{pak97,gaum1}. In Ref.
\cite{gaum2} author and others studied the isospin effects in
E$_{bal}$ at all the colliding geometries. A very few studies have
been carried out to study other related phenomena at E$_{bal}$ of
the neutron-rich systems. An important quantity which reflects the
dynamics in a heavy-ion collision is the density and temperature
reached in a reaction. In the present paper, we study the density
and temperature reached in heavy-ion reactions of neutron-rich
matter at E$_{bal}$ using isospin-dependent quantum molecular
dynamics (IQMD) model \cite{hart98}.

\begin{figure}[!t] \centering
 \vskip -1cm
\includegraphics[width=12cm]{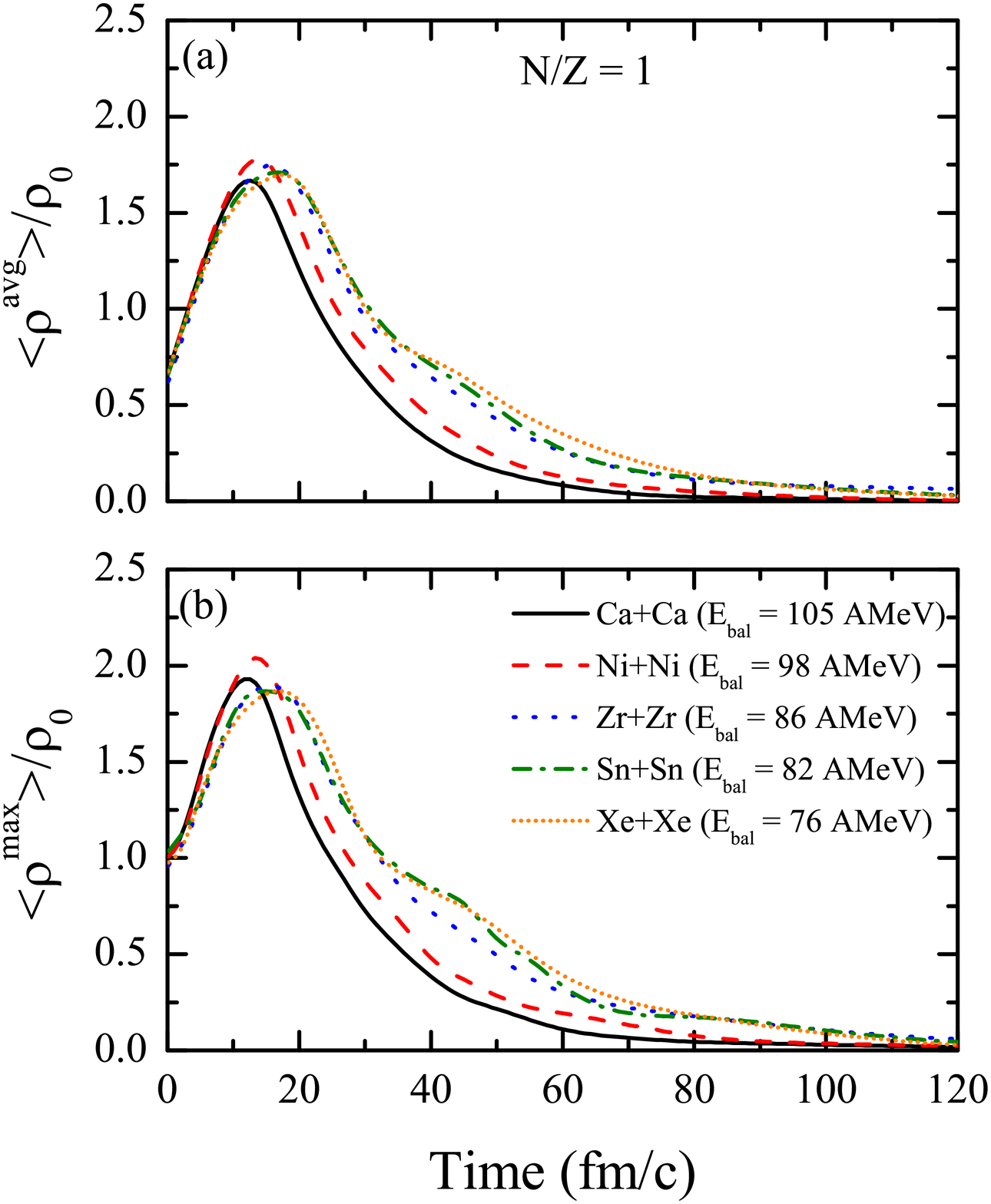}
\caption{(Color online) The time evolution of average density
(upper panel) and maximum density (lower panel) for systems having
N/Z = 1.0. Lines are explained in the text.}\label{fig1}
\end{figure}

\begin{figure}[!t] \centering
\vskip -1.cm
\includegraphics[angle=0,height=15cm,width=12.cm]{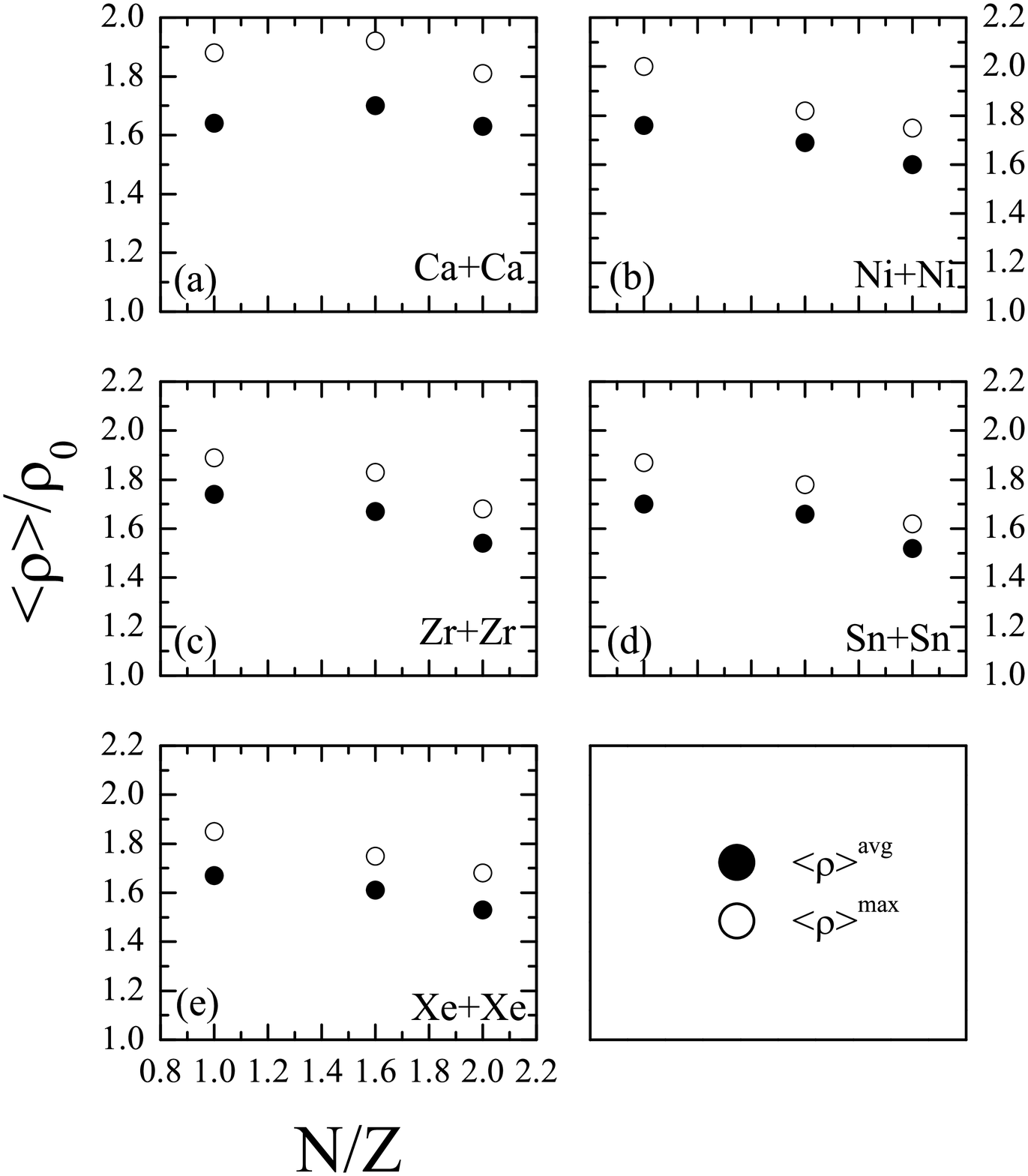}
\vskip -0.8cm \caption{ The N/Z dependence of maximal value of
average (solid circles) and maximum density (open circles) for
different N/Z ratios.}\label{fig3}
\end{figure}

\section*{Results and discussion}
We simulate the reactions of Ca+Ca, Ni+Ni, Zr+Zr, Sn+Sn, and Xe+Xe
series having N/Z = 1.0, 1.6 and 2.0. In particular, we simulate
the reactions of $^{40}$Ca+$^{40}$Ca (105), $^{52}$Ca+$^{52}$Ca
(85), $^{60}$Ca+$^{60}$Ca (73); $^{58}$Ni+$^{58}$Ni (98),
$^{72}$Ni+$^{72}$Ni (82), $^{84}$Ni+$^{84}$Ni (72);
$^{81}$Zr+$^{81}$Zr (86), $^{104}$Zr+$^{104}$Zr (74),
$^{120}$Zr+$^{120}$Zr (67); $^{100}$Sn+$^{100}$Sn (82),
$^{129}$Sn+$^{129}$Sn (72), $^{150}$Sn+$^{150}$Sn (64) and
$^{110}$Xe+$^{110}$Xe (76), $^{140}$Xe+$^{140}$Xe (68) and
$^{162}$Xe+$^{162}$Xe (61) at an impact parameter of
b/b$_{\textrm{max}}$ = 0.2-0.4  and at the incident energies equal
to
 balance energy. The values in the brackets represent
the balance energies for the systems. In fig. 1(a), we display the
time evolution of average density ($\rho^{avg}/\rho_{0}$) whereas
fig. 1(b) displays the time evolution of maximum density
($\rho^{max}/\rho_{0}$) for the systems having N/Z = 1.0, i.e, we
display the reactions of $^{40}$Ca+$^{40}$Ca, $^{58}$Ni+$^{58}$Ni,
$^{81}$Zr+$^{81}$Zr, $^{100}$Sn+$^{100}$Sn, and
$^{110}$Xe+$^{110}$Xe at energy equal to balance energy. From
figure, we find that maximal value of $\rho^{avg}/\rho_{0}$ is
higher for lighter systems as compared to the heavier ones.
Moreover, the density profile is more extended in heavier systems
indicating that the reaction finishes later in heavier systems.
This is because of the fact that the heavier reaction occurs at
low incident energy. Also the $\rho^{avg}/\rho_{0}$ and
$\rho^{max}/\rho_{0}$ are nearly same for heavier systems but
differ for lighter systems. Further, the maximum and average
densities are comparable for medium and heavy mass systems
indicating that the dense matter is formed widely and uniformly in
the central zone of the reaction. On the other hand, the
substantial difference in two densities for the lighter colliding
nuclei indicates the non-homogeneous nature of dense matter.
\par
In fig. 2 we display the N/Z dependence of maximal value of
$\rho^{avg}$ and $\rho^{max}$. Solid (open) symbols display the
results for  $\rho^{avg}$ ( $\rho^{max}$). From figure we see that
both  $\rho^{avg}$  and  $\rho^{avg}$  decreases slightly with N/Z
of the system for all the system masses. A slight exception to
this is there for the lighter mass of Ca+Ca \cite{gaum3}.

\section*{Acknowledgments}
 This work has been supported by a grant from Centre of Scientific
and Industrial Research (CSIR), Govt. of India.

\end{document}